\documentclass[fleqn,usenatbib]{mnras}

\usepackage{newtxtext,newtxmath}

\usepackage[T1]{fontenc}
\usepackage{ae,aecompl}

\usepackage{graphicx}	
\usepackage{amsmath}	
\usepackage{amssymb}	

\newcommand{\gaia}{{\it Gaia}}

\newcommand{\TESS}{{\it TESS}}

\newcommand{\kms}{km\,s$^{-1}$}

\newcommand{\masy}{mas\,y$^{-1}$}
\newcommand{\mpl}{\mbox{M$_{p}$}}
\newcommand{\rpl}{\mbox{R$_{p}$}}
\newcommand{\mstar}{\mbox{M$_{s}$}}
\newcommand{\rstar}{\mbox{R$_{s}$}}
\newcommand{\mjup}{\mbox{M$_{J}$}}
\newcommand{\rjup}{\mbox{R$_{J}$}}
\newcommand{\msun}{\mbox{$M_{\odot}$}}
\newcommand{\rsun}{\mbox{$R_{\odot}$}}
\newcommand{\rearth}{R$_{\oplus}$}
\newcommand{\gccc}{g\,cm$^{-3}$}

\newcommand{\teff}{$T_{\rm eff}$}
\newcommand{\logg}{$\log g$}
\newcommand{\av}{A$_{\rm V}$}
\newcommand{\lstar}{L$_{s}$}
\newcommand{\lsun}{L$_{\odot}$}

\title[NGTS-6b: A new ultra short period hot Jupiter discovery]{NGTS-6b: An Ultra Short Period Hot-Jupiter Orbiting an Old K Dwarf}

\author[J. I. Vines]{
\parbox{\textwidth}{
Jose I. Vines,$^{1}$\thanks{E-mail: jose.vines@ug.uchile.cl}
James S. Jenkins,$^{1,2}$
Jack~S.~Acton,$^{6}$
Joshua Briegal,$^{9}$
Daniel~Bayliss,$^{4}$
Fran\c{c}ois Bouchy,$^{8}$
Claudia~Belardi,$^{6}$
Edward M. Bryant,$^{4, 5}$
Matthew~R.~Burleigh,$^{6}$
Juan~Cabrera,$^{3}$
Sarah~L.~Casewell,$^{6}$
Alexander Chaushev,$^{10}$
Benjamin~F.~Cooke,$^{4,5}$
Szil\'ard~Csizmadia,$^{3}$
Philipp~Eigm\"uller,$^{3}$
Anders~Erikson,$^{3}$
Emma~Foxell,$^{4}$
Samuel Gill,$^{4,5}$
Edward~Gillen,$^{9,\dagger}$
Michael~R.~Goad,$^{6}$
James A. G. Jackman,$^{4,5}$
George~W.~King,$^{4,5}$
Tom Louden,$^{4,5}$
James~McCormac,$^{4,5}$
Maximiliano~Moyano,$^{12}$
Louise~D.~Nielsen,$^{8}$
Don~Pollacco,$^{4,5}$
Didier~Queloz,$^{9}$
Heike~Rauer,$^{3,10,11}$
Liam Raynard,$^{6}$
Alexis~M.~S.~Smith,$^{3}$
Maritza~G.~Soto,$^{13}$
Rosanna~H.~Tilbrook,$^{6}$
Ruth~Titz-Weider,$^{3}$
Oliver~Turner,$^{8}$
St\'{e}phane~Udry,$^{8}$
Simon.~R.~Walker,$^{4}$
Christopher~A.~Watson,$^{7}$
Richard~G.~West,$^{4,5}$
Peter~J.~Wheatley$^{4,5}$
}
\\
$^{1}$Departamento de Astronom\'ia, Universidad de Chile, Casilla 36-D, Santiago, Chile\\
$^{2}$ Centro de Astrof\'isica y Tecnolog\'ias Afines (CATA), Casilla 36-D, Santiago, Chile.\\
$^{3}$Institute of Planetary Research, German Aerospace Center, Rutherfordstrasse 2, 12489 Berlin, Germany\\
$^{4}$Dept.\ of Physics, University of Warwick, Gibbet Hill Road, Coventry CV4 7AL, UK\\
$^{5}$Centre for Exoplanets and Habitability, University of Warwick, Gibbet Hill Road, Coventry CV4 7AL, UK\\
$^{6}$Department of Physics and Astronomy, University of Leicester, University Road, Leicester, LE1 7RH, UK\\
$^{7}$Astrophysics Research Centre, School of Mathematics and Physics, Queen's University Belfast, BT7 1NN Belfast, UK\\
$^{8}$Observatoire de Gen{\`e}ve, Universit{\'e} de Gen{\`e}ve, 51 Ch. des Maillettes, 1290 Sauverny, Switzerland\\
$^{9}$Astrophysics Group, Cavendish Laboratory, J.J. Thomson Avenue, Cambridge CB3 0HE, UK\\
$^{10}$Center for Astronomy and Astrophysics, TU Berlin, Hardenbergstr. 36, D-10623 Berlin, Germany\\
$^{11}$Institute of Geological Sciences, FU Berlin, Malteserstr. 74-100, D-12249 Berlin, Germany\\
$^{12}$Instituto de Astronom\'ia, Universidad Cat\'olica del Norte, Angamos 0610, 1270709, Antofagasta, Chile\\
$^{13}$School of Physics and Astronomy, Queen Mary University, 327 Mile End Road, London E1 4NS, UK\\
$^{\dagger}$ Winton Fellow
}

\date{Accepted XXX. Received YYY; in original form ZZZ}

\pubyear{2019}

\begin{document}
\label{firstpage}
\pagerange{\pageref{firstpage}--\pageref{lastpage}}
\maketitle

\begin{abstract}

We report the discovery of a new ultra-short period hot Jupiter from the Next Generation Transit Survey.  NGTS-6b orbits its star with a period of 21.17~h, and has a mass and radius of $1.330^{+0.024}_{-0.028}$\mjup\, and $1.271^{+0.197}_{-0.188}$\rjup\, respectively, returning a planetary bulk density of 0.711$^{+0.214}_{-0.136}$~g~cm$^{-3}$.  Conforming to the currently known small population of ultra-short period hot Jupiters, the planet appears to orbit a metal-rich star ([Fe/H]$=+0.11\pm0.09$~dex).  Photoevaporation models suggest the planet should have lost 5\% of its gaseous atmosphere over the course of the 9.6~Gyrs of evolution of the system. NGTS-6b adds to the small, but growing list of ultra-short period gas giant planets, and will help us to understand the dominant formation and evolutionary mechanisms that govern this population.

\end{abstract}




\section{Introduction}

Over the last few years, ultra-short period (USP) planets have emerged as an important sub-population of planets, characterized solely by their proximity to the host star ($P_\text{orb} < 1$~day).  The majority of the population have been detected by space-based instruments, particularly CoRoT \citep{COROT} and Kepler \citep{KEPLER}, due to the tendency of the population to heavily favour small physical sizes and masses, and therefore large densities \citep{2011Natur.480..496C, 2013Natur.503..377P, 2017A&A...608A..93G, 2018NatAs...2..393S, 2018RNAAS...2c.172C, 2019arXiv190307694E}.

Ground-based radial velocity programs have found it difficult to detect these systems, since large-scale and high-cadence data sets are scarce, whereas the operational mechanics of photometric surveys allow for the detection of these systems.
Also, the formation mechanism seems to favour small planets, and therefore both the radial-velocity and transit signals are also small, making them harder to detect.  However, on the plus side, other biases work in favour of these methods, since the radial-velocity amplitude for a given star-planet systems increases with decreasing orbiting period, and the probability of transits rises, as well as the frequency of transits. 
In fact, although small USP super-Earths ($R_\text{p} \le 2R_{\oplus}$) are more common than larger planets by a factor of 5 \citep{USP_WINN}, a number have been detected from ground-based photometric surveys and confirmed by radial-velocity measurements, with the majority of this small sample being hot Jupiters \citep[HJs;][]{2009ApJ...707..167S, 2015MNRAS.447..711S, 2016AJ....152..127P, 2017AJ....153...97O}.  The population in between these two extremes of hot super-Earths and HJs have remained fairly elusive.

Models of the population of USP planets employ either photoevaporation or Roche Lobe overflow of a migrating more massive planet, which strips the planet of its gaseous envelope \citep{2014ApJ...793L...3V, 2016CeMDA.126..227J}.  The migration occurs either as disk migration \citep{2007ApJ...660..823M, 2014MNRAS.444.1738T} or by dynamical interactions \citep{2007ApJ...669.1298F}. In situ formation has also been invoked to describe the population \citep{2013MNRAS.431.3444C}.  The first USP HJ, WASP-19b \citep{WASP19}, has led to a deeper understanding of these models since a lack of explanation of why it has not lost the majority of its gaseous envelope provides strong constraints on the history of its dynamical evolution \citep{Essick_2015}.  Subsequent and forthcoming discoveries of USP HJ planets are providing critical constraints on the formation and evolution of close-in planets.

The Next Generation Transit Survey (NGTS; \citealt{NGTS_2012}; \citealt{NGTS_2013}; \citealt{NGTS_2017}; \citealt{NGTS_2018}) has now been fully operational for over two years, announcing the discovery of 5 new planets (\citealt{NGTS_1b, NGTS_2b, NGTS_3b, NGTS_4b, NGTS_5b}), which include a dense sub-Neptune (NGTS-4b), a sub-Jovian planet (NGTS-5b), a giant planet transiting an M dwarf star (NGTS-1b), and some new HJs (NGTS-2b, NGTS-3Ab).  The dense sampling of NGTS fields over long observing seasons, combined with the high precision of individual images ($\sim$0.001~magnitudes for a 1 hour baseline on stars brighter than I=14 during dark time or I=13 in full moon nights), allows the detection of not only smaller transiting planets, but also those with very short periods.  In this work, we report the discovery of a new USP HJ, NGTS-6b, orbiting the star NGTS-6.  The paper is organised as follows; in $\S$~\ref{sec:obs} we describe the NGTS observations that led to the discovery, with follow-up photometry from SAAO discussed in $\S$~\ref{sec:phot_follow} and the follow-up spectroscopy from FEROS and CORALIE discussed in $\S$~\ref{sec:spect_follow}. We analyse the nature of the star in $\S$~\ref{sec:stellarParams} and discuss the modeling in $\S$~\ref{sec:modeling}. Our conclusions are highlighted in $\S$~\ref{sec:close}.

\section{Observations}
\label{sec:obs}

\subsection{NGTS Photometry}

The NGTS is a ground-based wide-field transit survey located at the ESO Paranal Observatory in Chile, monitoring stars with $I < 16$ \citep{NGTS_2018}. It obtains full-frame images from twelve independent telescopes, each with a field of view of 8 square degrees, at 13\,s cadence. The telescopes have apertures of 20\,cm and observe at a bandpass of 520-890\,nm. They have a total instantaneous field of view of 96\,square degrees. A total of 213,549 10s exposures of the field containing NGTS-6 were taken between the nights of 16th of August 2017 and the 23rd of March 2018.

\begin{figure}
	\includegraphics[width=\columnwidth]{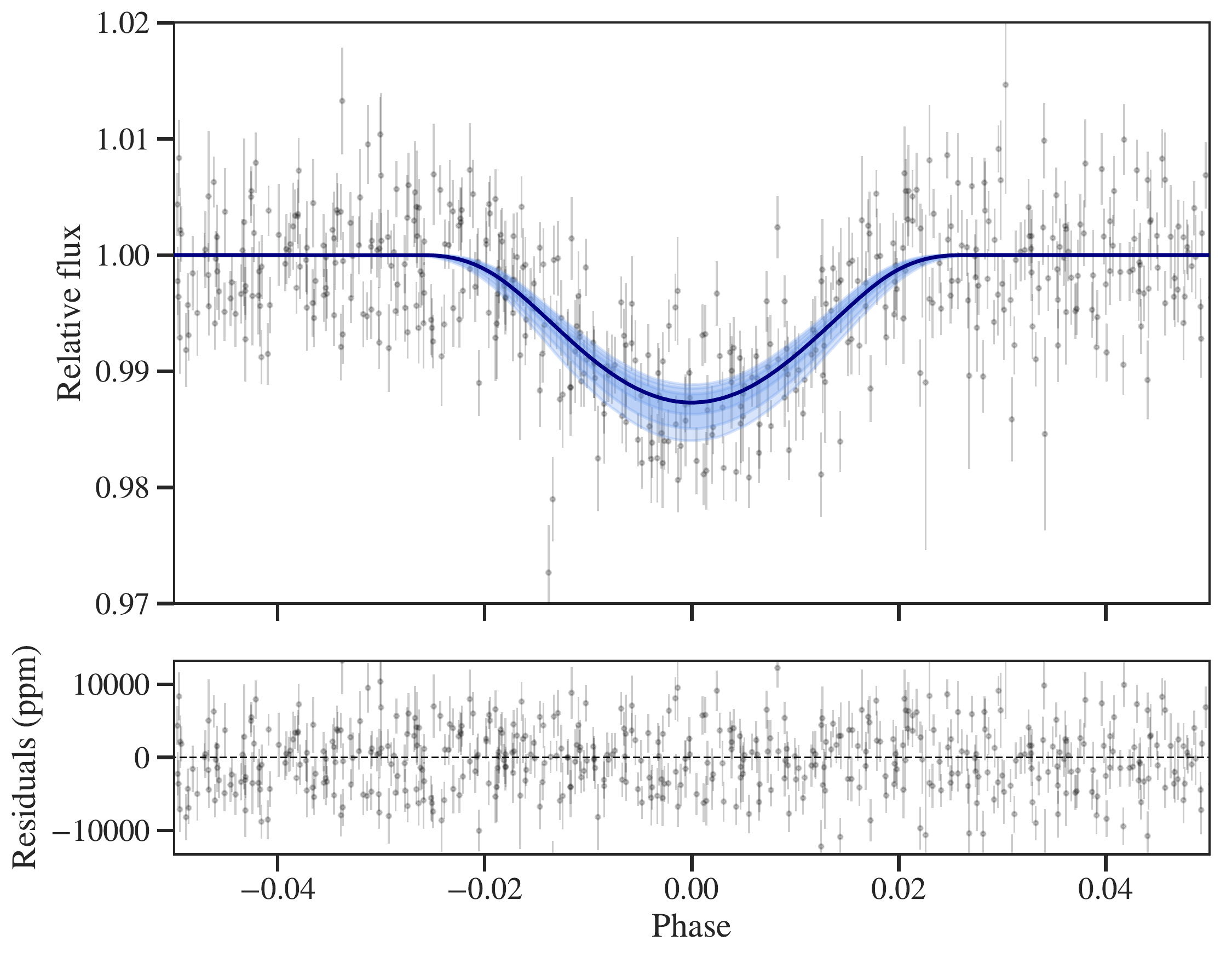}
    \caption{Light curve of NGTS photometry for NGTS-6 phase folded to the planets orbital period. The grey circles show the photometry observations binned to 10 minute cadence and the figure is zoomed to highlight the transit. The solid blue line and blue shaded regions represent the median, $1,2,3\sigma$ confidence levels,  respectively, of the best posterior model. Bottom: The residuals of the fit in ppm.}
    \label{fig:NGTSLC}
\end{figure}

The NGTS data were processed using the NGTS pipeline \citep{NGTS_2018}. Aperture photometry based on the CASUtools \citep{CASUtools} software package was used to generate the light curve using an aperture of 3 pixels, with 5 arcseconds per pixel. To remove the most dominant systematic effects the SysRem algorithm \citep{SysRem} has been utilized. Finally, using ORION, our own implementation of the BLS algorithm \citep{BLS}, a transit signal with a period of 0.882055~d for the star NGTS-6b was discovered.  We show the transit in the NGTS phase folded light curve with the corresponding model and confidence region (see Section \ref{sec:global_fit} for details) in Figure \ref{fig:NGTSLC}.

\subsection{TESS Photometry}
\begin{figure}
	\includegraphics[width=\columnwidth]{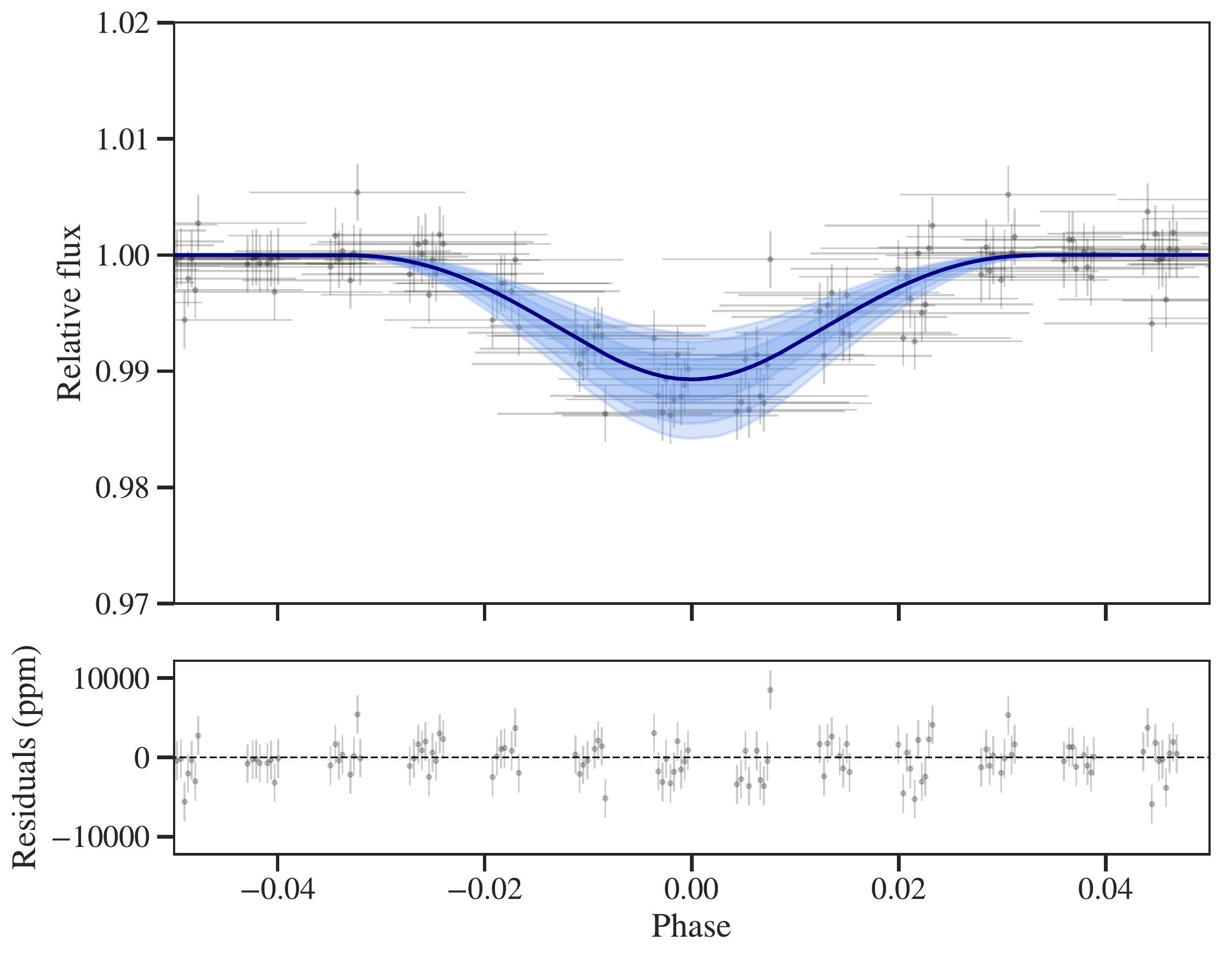}
    \caption{Phase folded TESS photometry. Horizontal errorbars show the 30 minute cadence of the observations. The solid blue line is the best fit for the photometry and the blue shaded regions represent the $1,2,3\sigma$ confidence levels. Bottom: The residuals of the fit in ppm.}
    \label{fig:TESSLC}
\end{figure}

The Transiting Exoplanet Survey Satellite (TESS) is a NASA-sponsored Astrophysics Explorer-class mission that is performing a wide-field survey to search for planets transiting bright stars \citep{TESS}.  It has four $24\times24\degr$ field of view cameras with four 2k$\times$2k CCDs each, with a pixel scale of 21 arcseconds per pixel and a bandpass of 600-1000\,nm. Using the TESSCut tool \footnote{https://mast.stsci.edu/tesscut} we checked for available data in the TESS full frame images. 
NTGS-6b was observed by TESS in Sector 5 using CCD 2 of Camera 2. Between November 15th and December 11th 2018 1196 images with a typical cadence of 30 minutes were obtained. Using a small 2x2 pixel aperture to minimize contamination, we performed aperture photometry on the target. Long term trends visible in the data were removed using a moving median filter. The single transit events are directly visible in the TESS data, although with the long cadence and short transit duration, not well sampled.
In Fig.~\ref{fig:TESSLC} the phase folded TESS light curve is shown.

\subsection{SAAO Photometric Follow Up}
\label{sec:phot_follow}

Three transit light curves were obtained with the 1.0-m Elizabeth telescope at the South African Astronomical Observatory (SAAO) and one of the SHOC frame-transfer CCD cameras, ``{\em SHOC'n'awe}" \citep{SHOC}. The transits were collected on 2018 October 7 in $V$ band ($240 \times 60$\,second exposures), and on 2018 November 14 and 15 in $I$ band ($470 \times 30$\,second and $340 \times 30$\,s exposures respectively. The scale of each pixel is 0.167 arcsec). The data were reduced with the local SAAO SHOC pipeline, which is driven by {\sc python} scripts running {\sc iraf} tasks ({\sc pyfits} and {\sc pyraf}), and incorporating the usual bias and flat-field calibrations. Aperture photometry was performed  using the Starlink package {\sc autophotom}. 

Differential photometry was performed on each light curve using 2 reference stars and altering the size of the aperture to reflect the sky conditions (4px for the $V$ band light curve, 5px for the $I$ band light curve on 2018 Nov 14 and 3px for the light curve obtained on the following night when the conditions were considerably better).  

\begin{figure}

  \centering
  \includegraphics[width=\columnwidth]{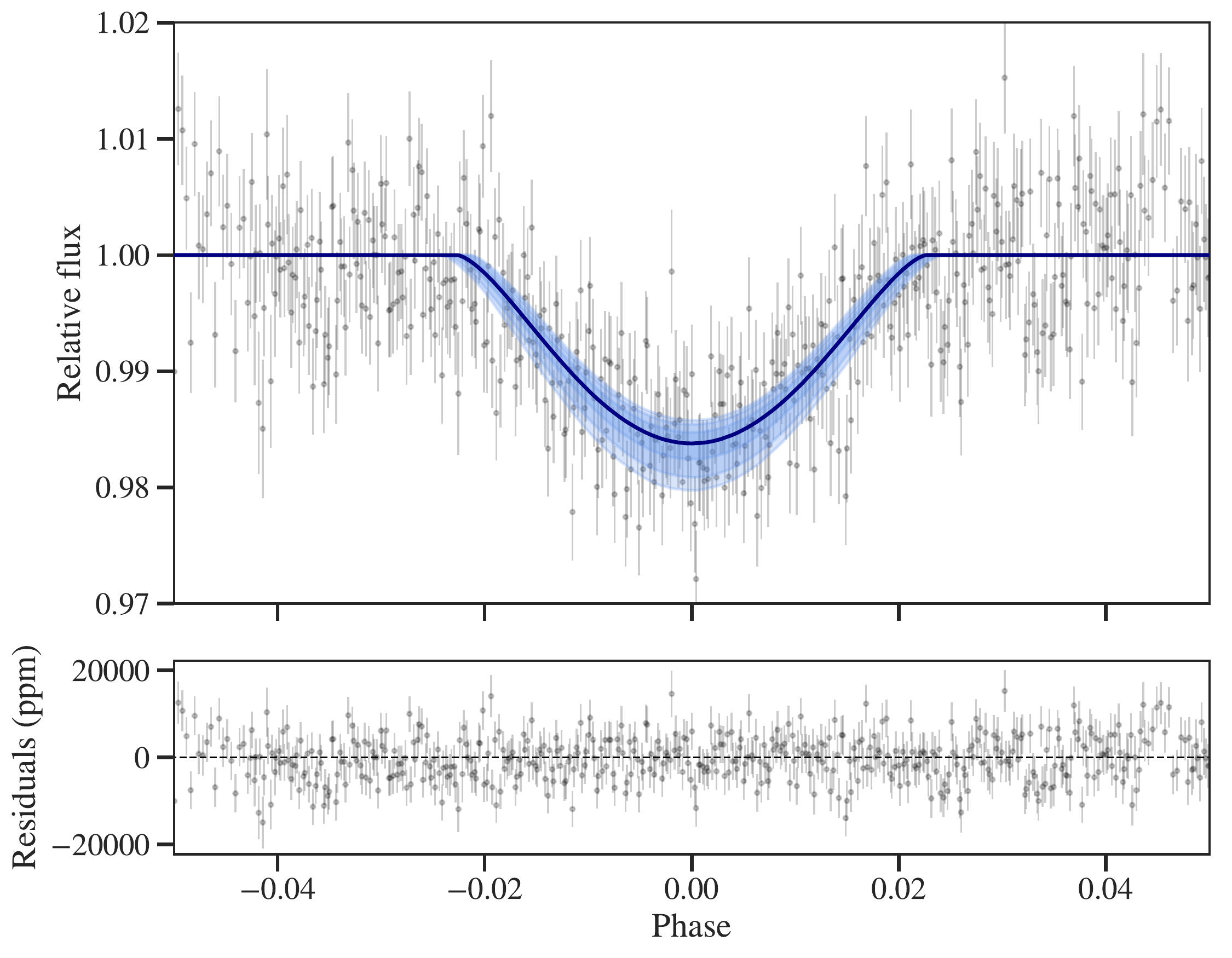}

  \centering
  \includegraphics[width=\columnwidth]{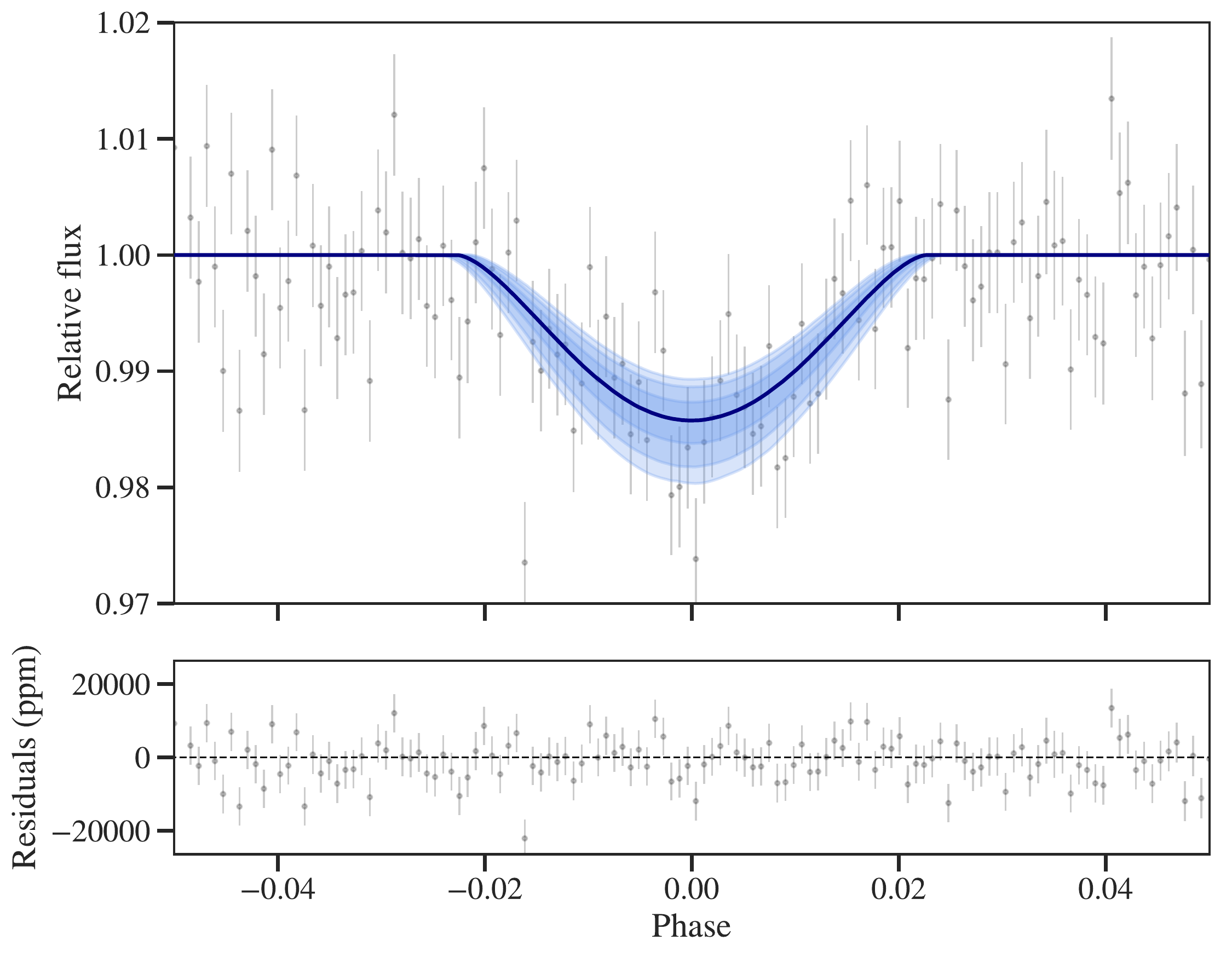}

\caption{Top: Phase folded SAAO photometry in the $I$ and $V$ band respectively for (a) and (b). The solid blue line is the best fit for the photometry and the blue shaded regions represent the $1,2,3\sigma$ confidence levels. Bottom: The residuals of the fit in ppm.}
\label{fig:SAAOLC}
\end{figure}

We show phase folded light curves of the three SAAO transit events in Figure \ref{fig:SAAOLC} with their respective models and confidence regions.  The complete light curve for all instruments is shown in Table \ref{tab:LC}

\begin{table}
	\centering
	\caption{Photometry of NGTS, SAAO I, V and TESS for NGTS-6. The full table is available in a machine-readable format from the online journal. A portion is shown here for guidance.}
	\label{tab:LC}
\begin{tabular}{cccc}

    BJD        &  Relative Flux& Relative Flux &     INST \\
(-2,450,000)   &               &       error &    \\ \hline
    \ldots &       \ldots &       \ldots &       \ldots \\
    8200.55123 &       1.0222 &       0.0169 &  NGTS \\
    8200.55138 &       0.9660 &       0.0168 &  NGTS \\
    8200.55153 &       0.9859 &       0.0169 &  NGTS \\
    8200.55168 &       1.0206 &       0.0171 &  NGTS \\
    8200.55183 &       0.9861 &       0.0169 &  NGTS \\
    8399.47400 &       1.0049 &       0.0044 & SAAOV \\
    8399.47469 &       1.0077 &       0.0044 & SAAOV \\
    8399.47539 &       0.9974 &       0.0044 & SAAOV \\
    8399.47608 &       0.9982 &       0.0044 & SAAOV \\
    8399.47678 &       1.0014 &       0.0043 & SAAOV  \\
    \ldots & \ldots & \ldots & \ldots \\ \hline
	\end{tabular}
\end{table}

\subsection{Spectroscopic Follow Up}
\label{sec:spect_follow}

We obtained multi-epoch spectroscopy for NGTS-6\ with two different fiber-fed high precision  \'echelle spectrographs: CORALIE and FEROS. Both are located at the ESO La Silla Observatory in Chile. CORALIE is mounted on the 1.2-m Leonard Euler telescope and has a spectral resolution of R = 60,000 \citep{CORALIE}. FEROS is mounted on the 2.2-m MPG/ESO telescope with a spectral resolution of $R = 48,000$ \citep{FEROS}.  The spectral observations were taken between October 23 2018 and January 8 2019 for CORALIE and December 23 2018 and January 2 2019 for FEROS.

The CORALIE data were reduced using the standard data reduction pipeline \citep{CORALIE} and the radial velocities were calculated by cross-correlation with a binary G2 mask. The first 20 orders of the spectrum were discarded in the cross-correlation analysis as they contain little signal. All the CORALIE spectra were also stacked to make a high signal-to-noise spectrum for spectral analysis as presented in Sect. \ref{sec:stellarParams}

FEROS data were reduced with the CERES pipeline \citep{CERES_PIPELINE}. CERES also calculates the cross-correlation function (CCF) using the reduced FEROS spectra and a binary G2 mask for each epoch and afterwards, depending on moonlight contamination, a single or double Gaussian is then fitted to find the radial velocity (depending on moonlight contamination). In the cases where the single or double Gaussian fits were unsatisfactory, a 4th order spline was fitted to find the radial velocity instead. The radial velocities from CORALIE and FEROS are shown in Table \ref{tab:rvs} along with the uncertainties and bisector velocity span. We present a total of 21 radial velocity data points, which constrain the orbit of the planet, 11 of which were taken with CORALIE and 10 with FEROS.

\begin{figure}
	\includegraphics[width=\columnwidth]{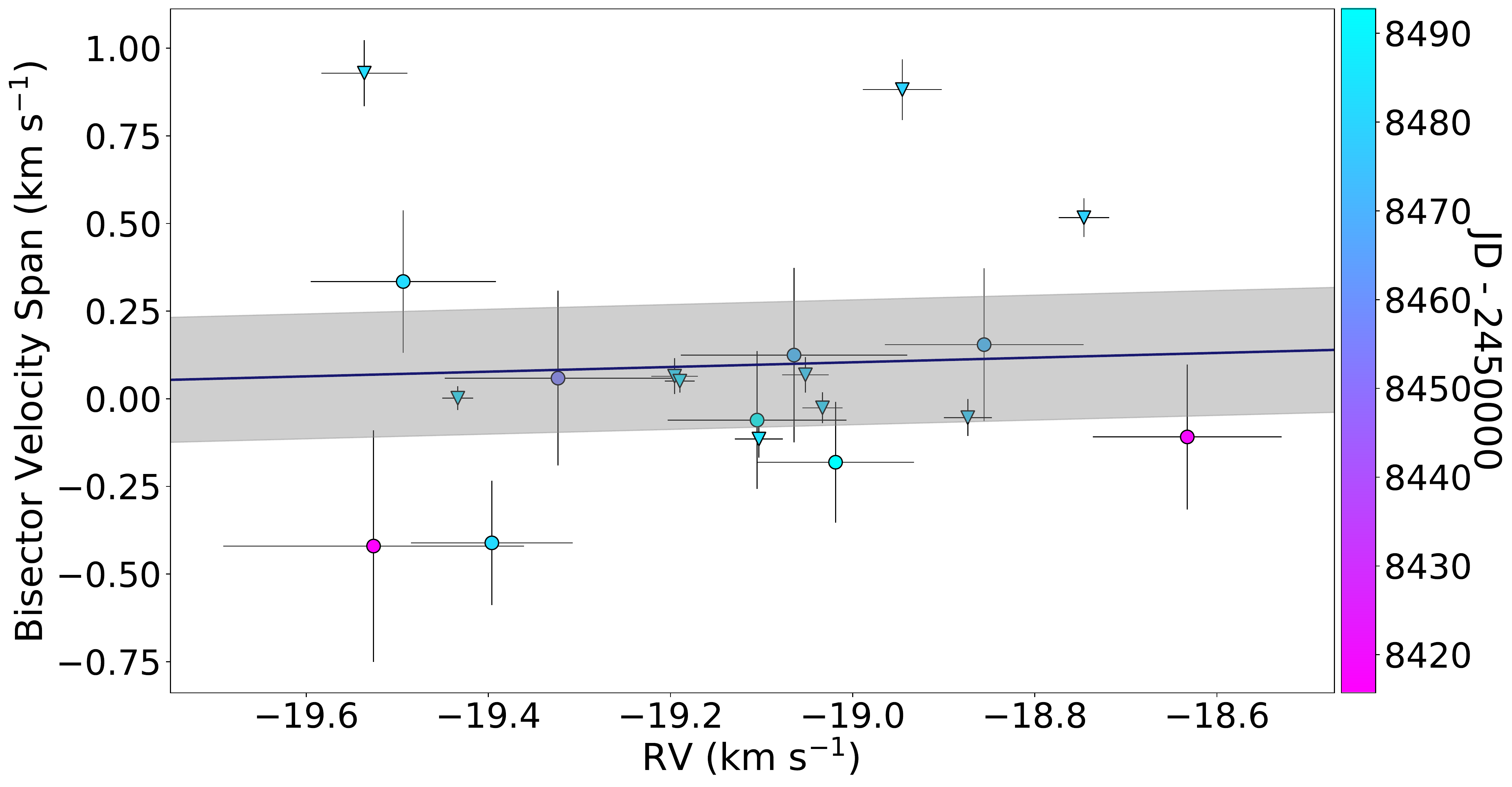}
    \caption{Bisector velocity span over radial velocity measurements color coded by observation time. Circles are CORALIE and upside-down triangles are FEROS datapoints. The blue solid line is a linear fit and the shaded region show the $2\sigma$ confidence region. No correlation is detected.}
    \label{fig:bisector}
\end{figure}

As a first check we searched for a correlation between the radial-velocity data and the bisector velocity span, which was calculated using the CCFs that were constructed to measure the velocity in the first place (see for example \citealt{2009A&A...495..959B}).  Any correlation between radial-velocity measurements and the bisector velocity span would cast doubts on the validity of the planet interpretation of the radial-velocity variation \citep{BIS_RV_CORR}. The bisector and radial-velocities are shown in Figure \ref{fig:bisector} along with a linear fit and a $2\sigma$ confidence region, with no correlation detected. We calculated the Pearson r coefficient to be $0.05$, which reaffirms our claim of no significant linear correlation being present.

\begin{table}
	\centering
	\caption{CORALIE and FEROS Radial Velocities for NGTS-6}
	\label{tab:rvs}
\begin{tabular}{ccccc}
    BJD &             RV &  $\sigma_{RV}$ &            BIS &     INST \\
  (-2,450,000) &  (km s$^{-1}$) &  (km s$^{-1}$) &  (km s$^{-1}$) &  \\ \hline
       8415.71 &       -19.526 &       0.165 &      -0.420 &  CORALIE \\
       8418.74 &       -18.633 &       0.104 &      -0.109 &  CORALIE \\
       8454.65 &       -19.324 &       0.125 &       0.059 &  CORALIE \\
       8472.59 &       -18.856 &       0.109 &       0.154 &  CORALIE \\
       8472.75 &       -19.064 &       0.124 &       0.125 &  CORALIE \\
       8475.69 &       -19.333 &       0.102 &       ----- &  CORALIE \\
       8475.79 &       -19.264 &       0.110 &       ----- &  CORALIE \\
       8481.71 &       -19.396 &       0.089 &      -0.411 &  CORALIE \\
       8481.81 &       -19.494 &       0.102 &       0.335 &  CORALIE \\
       8492.69 &       -19.019 &       0.086 &      -0.181 &  CORALIE \\
       8492.73 &       -19.105 &       0.098 &      -0.061 &  CORALIE \\
       8481.85 &       -19.536 &       0.047 &       0.929 &    FEROS \\
       8480.75 &       -19.196 &       0.026 &       0.064 &    FEROS \\
       8480.73 &       -19.033 &       0.022 &      -0.026 &    FEROS \\
       8484.71 &       -19.189 &       0.016 &       0.050 &    FEROS \\
       8478.79 &       -18.746 &       0.028 &       0.517 &    FEROS \\
       8478.60 &       -19.052 &       0.025 &       0.068 &    FEROS \\
       8478.81 &       -18.946 &       0.043 &       0.882 &    FEROS \\
       8478.62 &       -18.874 &       0.026 &      -0.054 &    FEROS \\
       8483.84 &       -19.103 &       0.026 &      -0.115 &    FEROS \\
       8483.67 &       -19.434 &       0.017 &       0.002 &    FEROS \\ \hline
	\end{tabular}
\end{table}

\section{Stellar Parameters} \label{sec:stellarParams}
Given the host star is relatively faint ($V$ =  14.087), the high-resolution echelle spectra used for the calculation of the radial velocity measurements have SNR too low for the accurate measurement of the equivalent of individual absorption lines, therefore it is not possible to obtain a constrained solution from the SPECIES code (\citealp{soto18}) for the stellar bulk parameters.  Therefore, we used two methods: the empirical SpecMatch tool \citep{SPECMATCH} with the combined CORALIE spectra, which given the observing conditions have a higher SNR than the FEROS spectra, and an SED fit of the star detailed in Sect. \ref{sec:sed}.
The output of the two employed methods are shown in Table \ref{tab:comp}

\begin{table}
	\centering
	\caption{Comparison of empirical SpecMatch and SED fitting outputs}
	\label{tab:comp}
\begin{tabular}{ccc}
     Parameter &              SpecMatch &   SED            \\ \hline
     \teff~(K) &  4409 $\pm$  70 &  4730$^{+44}_{-40}$ \\
     \logg &  4.63 $\pm$  0.12 &  4.7$^{+1.1}_{-0.7}$ \\
     \rstar~(\rsun) &  0.72 $\pm$  0.1 &  0.754 $\pm$  0.013 \\
     $\left[Fe/H\right]$ &  0.11 $\pm$  0.09 & ---- \\
    $(\dagger)$  \mstar~(\msun)&  0.72 $\pm$  0.08 &  0.767 $\pm$  0.025 \\
    $(\dagger)$  Age~(Gyr) &  9.61 $\pm$  0.17 &  9.77$^{+0.25}_{-0.54}$ \\
     distance~(pc) & ---- &  308 $\pm$  2 \\
    {\av} & ---- &  0.017 $\pm$  0.010 \\ \hline
    \multicolumn{3}{l} {In the case of SED fitting, the parameters with $(\dagger)$} \\
    \multicolumn{3}{l} {were calculated using the isochrones package.}
\end{tabular}
	
\end{table}

We finally adopt the results from the SED fitting routine for \teff, \logg, and \rstar, and the metallicity from the SpecMatch tool. We used these parameters to calculate the mass and age of the star, using the isochrones package \citep{2015ascl.soft03010M}. The projected rotational velocity, vsini, was estimated using the SPECIES code. We combined the individual spectra obtained with Coralie to obtain a high S/N spectrum from the target, and created synthetic absorption line profiles for four iron lines in the spectrum, using the ATLAS9 model atmospheres \citep{ATLAS9} and the atmospheric parameters previously obtained. We then broaden the absorption lines, by adjusting the rotational velocity, until they matched the observations. More details about this procedure can be found in \citet{soto18}. The obtained vsini is listed in Table \ref{tab:stellar}. Thus we conclude that NGTS-6 is an old K dwarf with an effective temperature of 4730$^{+44}_{-40}$~K, \logg\ of 4.7$^{+1.1}_{-0.7}$~dex, [Fe/H] of 0.11 $\pm$ 0.09~dex, radius of 0.754 $\pm$ 0.013~\rsun,\ mass of 0.767 $\pm$ 0.025~\msun\ and age of 9.77$^{+0.25}_{-0.54}$~Gyrs. We show NGTS-6 catalogue information and stellar parameters in Table \ref{tab:stellar}.

\subsection{SED Fitting and Dilution}
\label{sec:sed}
\begin{figure}
	\includegraphics[width=\columnwidth]{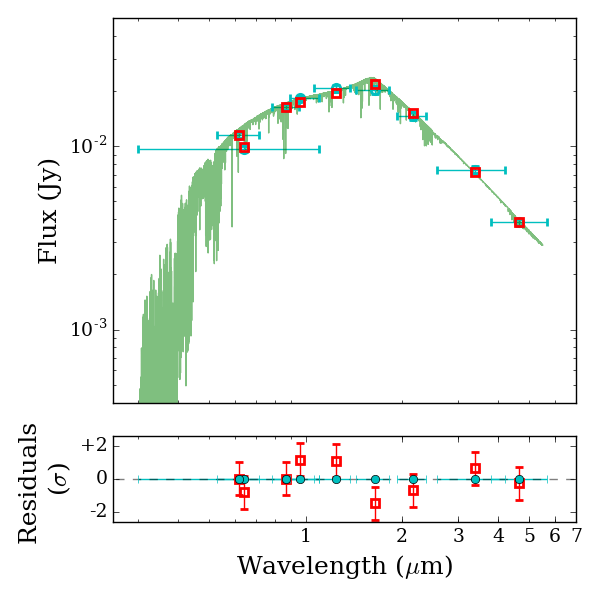}
    \caption{Top: The best fitting PHOENIX v2 SED model, obtained from fitting to unblended Pan-STARRS, \gaia, 2MASS and WISE photometry. The cyan and red points indicate the catalogue and synthetic photometry respectively. The horizontal error bars indicate the spectral coverage of each band. Bottom: Residuals of the synthetic photometry, normalised to the catalogue errors.}
    \label{fig:SED_fit}
\end{figure}

Using \gaia\ DR2 we identified a neighbouring source 5.4\arcsec\ away that could be contaminating our photometry. Therefore, in order to determine the level of dilution from this source we performed SED fitting of both stars using the PHOENIX v2 models \citep[][]{Husser13}. This was done following a method similar to \citet{Gillen17}, by firstly generating a grid of bandpass fluxes in \teff\ and \logg\ space. To overcome the issue of possible blending in catalogue photometry (due to the small separation of the two sources) we fit only to the Pan-STARRS, 2MASS, \gaia\ and WISE photometry for each source. We fit for the \teff, \logg, radius, distance, V band extinction \av\ and an uncertainty term $\sigma$ to account for underestimated catalogue uncertainties. We have used a Gaussian prior on the distance, constrained by the values calculated by \citet{BailerJones18} using \gaia\ DR2 data. We limit the V band extinction of each source to a maximum value of 0.032, taken from the  Galactic dust reddening maps of \citet{Schlafly11}. Before fitting we verified that neither source was flagged as extended in the catalogues used. Due to the larger size of the \TESS\ aperture, an extra source 15.9\arcsec\ was identified as possibly contributing light. Consequently, we also performed SED fitting on this source and included it for the \TESS\ dilution value only.
To sample the posterior parameter space for each source we used {\it emcee} \citep[][]{emcee} to create a Markov Chain Monte Carlo (MCMC) process for our fitting. In this process we used 100 walkers for 50,000 steps and discarded the first 10,000 as a burn in. The best fitting SED model for NGTS-6\ is shown in Fig.\,\ref{fig:SED_fit} and the values are given in Tab.\,\ref{tab:stellar}.

To estimate the level of dilution in each bandpass we convolved the SED model for each star with the specified filter, taking the ratio of the measured synthetic fluxes as the dilution value. In order to sample the full range of dilutions and thus provide an informative prior we draw our SED models directly from the posterior distribution for each star The calculated dilutions are D$_\text{NGTS} =$ 0.056 $\pm$ 0.002, D$_\text{SAAOV} =$ 0.025 $\pm$ 0.002, D$_\text{SAAOI} =$ 0.085 $\pm$ 0.003 and D$_\text{TESS} =$ 0.077 $\pm$ 0.003. With these results we generate priors for the dilution in our lightcurves, which are used in the transit fitting.

\begin{table}
	\centering
	\caption{Stellar Properties for NGTS-6}
	\begin{tabular}{lcc} 
	Property	&	Value		&Source\\
	\hline
    \multicolumn{3}{l}{Astrometric Properties}\\
    R.A.		&	$05^{\rmn{h}} 03^{\rmn{m}} 10\fs90$	&2MASS	\\
	Dec			&	$-30\degr 23\arcmin 57\farcs 6420$ 	&2MASS	\\
    2MASS I.D.	& 05031090-3023576	&2MASS	\\
    Gaia DR2 I.D. & 4875693023844840448 & {\em Gaia} \\
    TIC ID      & 1528696  &  TESS \\ 
    $\mu_{{\rm R.A.}}$ (\masy) & $-6.0 \pm 7.0$ & UCAC4 \\
	$\mu_{{\rm Dec.}}$ (\masy) & $-33.5 \pm 10.1$ & UCAC4 \\
	$\varpi$ (mas) & $3.215 \pm 0.015$ & {\em Gaia} \\
    \\
    \multicolumn{3}{l}{Photometric Properties}\\
	V (mag)		&$14.087 \pm 0.021$ 	&APASS\\
	B (mag)		&$15.171 \pm 0.014$		&APASS\\
	g (mag)		&$14.639 \pm 0.058$		&APASS\\
	r (mag)		&$13.703 \pm 0.032$		&APASS\\
	i (mag)		&$13.378 \pm 0.057$		&APASS\\
	$\mathrm{r_{P1}}$ (mag)	&$13.751 \pm 0.002$		&Pan-STARRS\\
	$\mathrm{z_{P1}}$ (mag) &$13.364 \pm 0.002$		&Pan-STARRS\\
	$\mathrm{y_{P1}}$ (mag) &$13.250 \pm 0.006$		&Pan-STARRS\\
    G (mag)		&$13.818 \pm 0.001$	&{\em Gaia}\\
    BP (mag)    &$14.401 \pm 0.003$   &{\em Gaia}\\
    RP (mag)    &$13.113 \pm 0.002$   &{\em Gaia}\\
    NGTS (mag)	&$13.460$	&This work\\
    TESS (mag)  &$13.070$   &TESS\\
    J (mag)		&$12.222 \pm 0.033$		&2MASS	\\
   	H (mag)		&$11.767 \pm 0.038$		&2MASS	\\
	K (mag)		&$11.650 \pm 0.032$		&2MASS	\\
    W1 (mag)	&$11.609 \pm 0.028$		&WISE	\\
    W2 (mag)	&$11.688 \pm 0.029$	    &WISE	\\
    \\
    \multicolumn{3}{l}{Derived Properties}\\

    T$_{\rm eff}$ (K)    &  4730$^{+44}_{-40}$     & SED fitting\\\vspace{2pt}
    log g       &		 4.7$^{+1.1}_{-0.7}$			& SED fitting\\\vspace{2pt}
    $\left[Fe/H\right]$	 & $0.11 \pm 0.09$		&CORALIE spectra\\
    vsini  (\kms)	 &  2.851 $\pm$ 0.431 & CORALIE spectra\\
    $\gamma_{CORALIE}$ (\kms) & $-19.137^{+0.018}_{-0.018}$&Global Modelling\\\vspace{2pt}
    $\gamma_{FEROS}$ (\kms) & $-19.142 \pm 0.010$  &Global Modelling\\\vspace{2pt}
    $\sigma_{CORALIE}$ (\kms) & $0.000^{+0.039}_{-0.041}$&Global Modelling\\\vspace{2pt}
    $\sigma_{FEROS}$ (\kms) & $0.036^{+0.023}_{-0.094}$ &Global Modelling\\\vspace{2pt}
    \lstar (\lsun) &  0.256 $\pm$ 0.009 &  SED fitting\\\vspace{2pt}
    \mstar (\msun) &  0.767 $\pm$ 0.025	    & SED fitting\\
    \rstar (\rsun) &  0.754 $\pm$ 0.013	        & SED fitting\\
    $\rho$ (\gccc) & $3.9304^{+0.0815}_{-0.0812}$          &Global Modeling\\\vspace{2pt}
    Age				&  9.77$^{+0.25}_{-0.54}$		& SED fitting\\
    Distance (pc)	&  $311.042 \pm 1.432$	&{\em Gaia}\\
    
	\hline
    \multicolumn{3}{l}{2MASS \citep{2MASS}; UCAC4 \citep{UCAC};}\\
    \multicolumn{3}{l}{APASS \citep{APASS}; WISE \citep{WISE};}\\
    \multicolumn{3}{l}{{\em Gaia} \citep{GAIA, GAIA_DR2}; TESS \citep{TESS_CAT}}\\
    \multicolumn{3}{l}{Pan-STARRS \citep{Tonry12,Chambers16}}
	\end{tabular}
    \label{tab:stellar}
\end{table}

\section{Data Modeling}
\label{sec:modeling}

\subsection{Pure Radial Velocity Modelling}

\begin{figure*}
	\includegraphics[width=2\columnwidth]{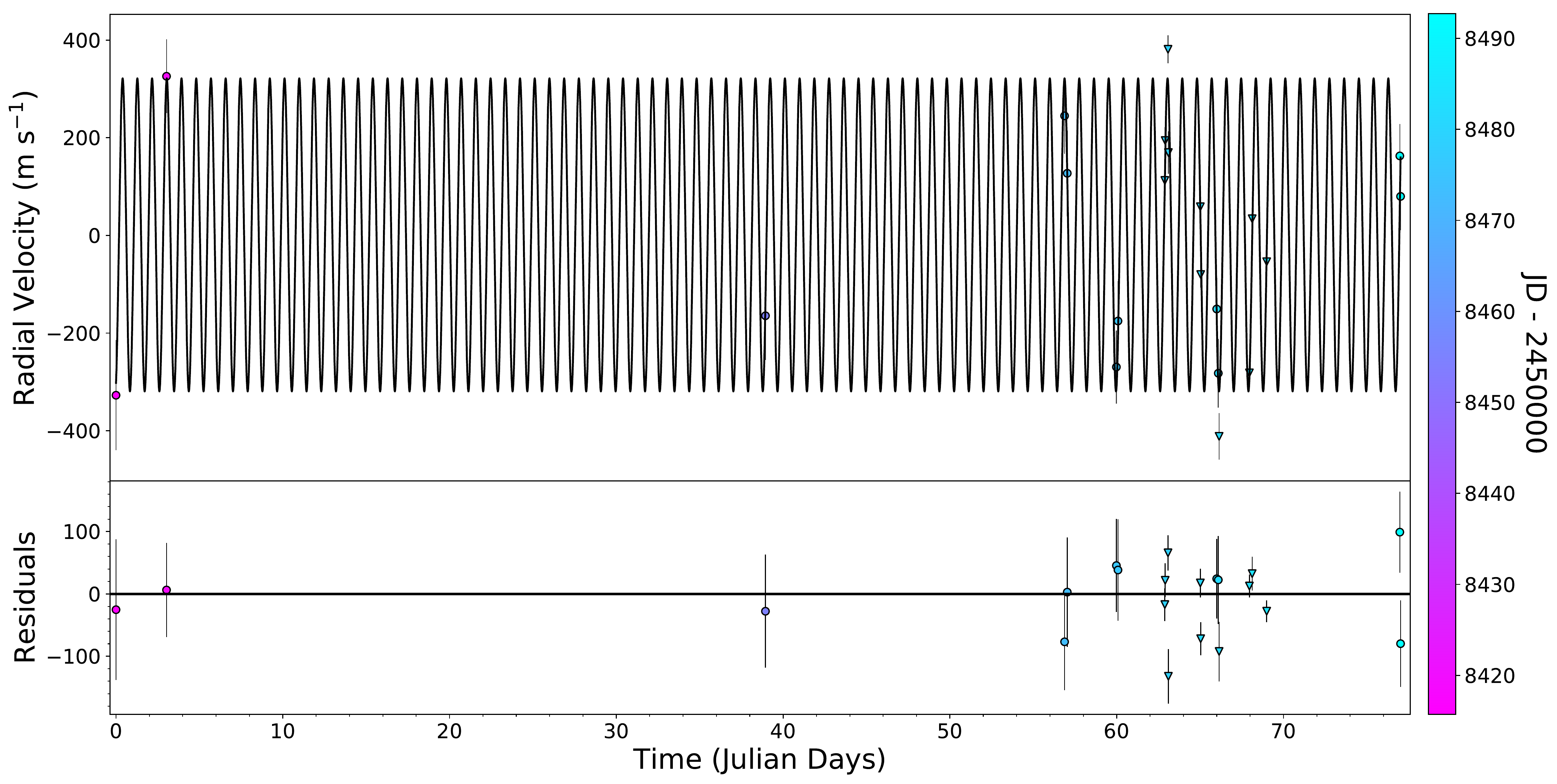}
    \caption{Top: The full timeseries of NGTS-6\, radial velocity observations, color coded by observation time. Circles are CORALIE datapoints and upside-down triangles are FEROS. The solid black line is the best Keplerian fit. Bottom: The residuals of the fit.}
    \label{fig:timeseries}
\end{figure*}

Firstly, a pure radial velocity search and model fit was made using the EMPEROR algorithm (\citealp{pena2019}).  EMPEROR is a public, python-based code that is designed to search for small signals in radial-velocity data using Bayesian modeling techniques and MCMC tools.  The algorithm allows for correlated noise models to be incorporated into the modeling, in particular moving averages of order selected by the user.  The code uses the affine-invariant {\it emcee} sampler in parallel tempering mode to efficiently sample highly multi-modal posteriors. 

\begin{figure*}
	\includegraphics[width=2\columnwidth]{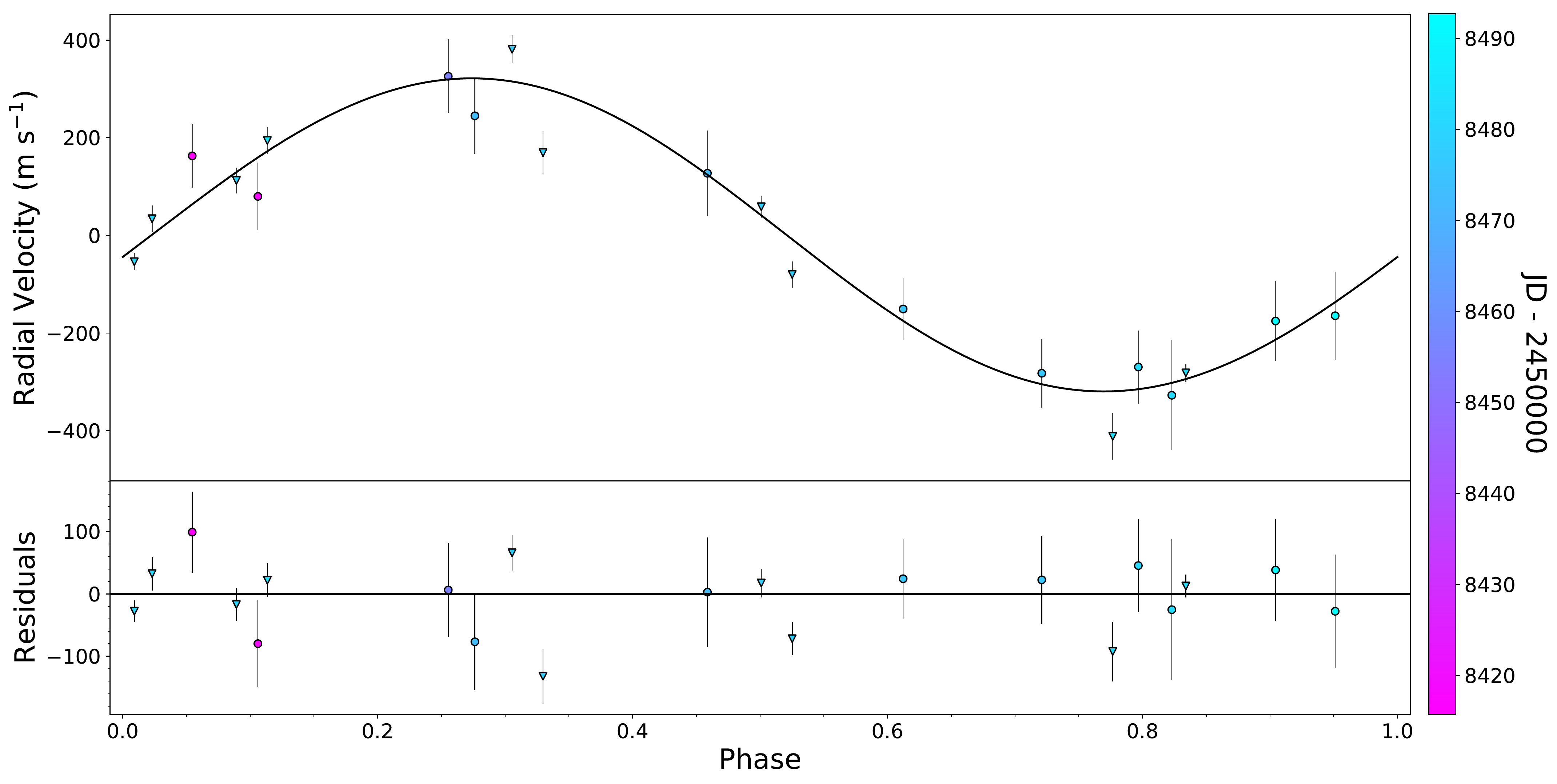}
    \caption{Top: NGTS-6\, radial velocity measurements and model in orbital phase, color coded by obervation times. Circles are CORALIE datapoints and upside-down triangles are FEROS. The solid black line is the fit to best orbital solution. Bottom: Residuals of the fit.}
    \label{fig:phasefold}
\end{figure*}

In order to first test if the signal was present in the data without the use of inputs from the photometry, as a test of signal independence, we employed six chains with different temperature values ($\beta = 1.0,\, 0.66,\, 0.44,\, 0.29,\, 0.19$ and $0.13$).  The chain length was set to 15,000 steps and each chain had 150 walkers in the ensemble, giving rise to a total chain length of 13.5 million steps.  A burn-in of 7,500 steps was also used.  A first-order moving average correlated noise model was used to model the high-frequency noise in the velocity data set, and the priors were set to be the standard priors as explained in the EMPEROR manuscript and on the GitHub page\footnote{https://github.com/ReddTea/astroEMPEROR}.  In automatic mode, EMPEROR detects the planet's orbital signature with a Bayes Factor value of 5, highly significant, in the combined FEROS+CORALIE data, confirming the existence of the planet. The best fit made by EMPEROR is shown in Figure \ref{fig:timeseries} and the phase folded curve in Figure \ref{fig:phasefold}.  No additional signal was detected. The best fitting model from EMPEROR with respective uncertainties were used as Gaussian priors to determine a global model for this system.

\subsection{Global Modeling}
\label{sec:global_fit}

For the global joint photometry and radial velocity modeling we used Juliet \citep{JULIET}. Juliet is a python tool capable of analysis of transits, radial velocities, or both. It allows the analysis of multiple photometry and radial velocity instruments at the same time using Nested Sampling, Importance Nested Sampling, and Dynamic Nested Sampling algorithms. For the transit models, Juliet uses BATMAN \citep{BATMAN}, which has flexible options, in particular for limb-darkening laws. The Keplerian signal model is provided by radvel \citep{RADVEL}. Finally for our Juliet run, given the high dimensionality of the model (29 free parameters between two radial velocity and four photometry instruments) we used Dynesty for Dynamic Nested Sampling as it has proven to be more efficient than regular Nested Sampling under these conditions.

\begin{figure}
	\includegraphics[width=\columnwidth]{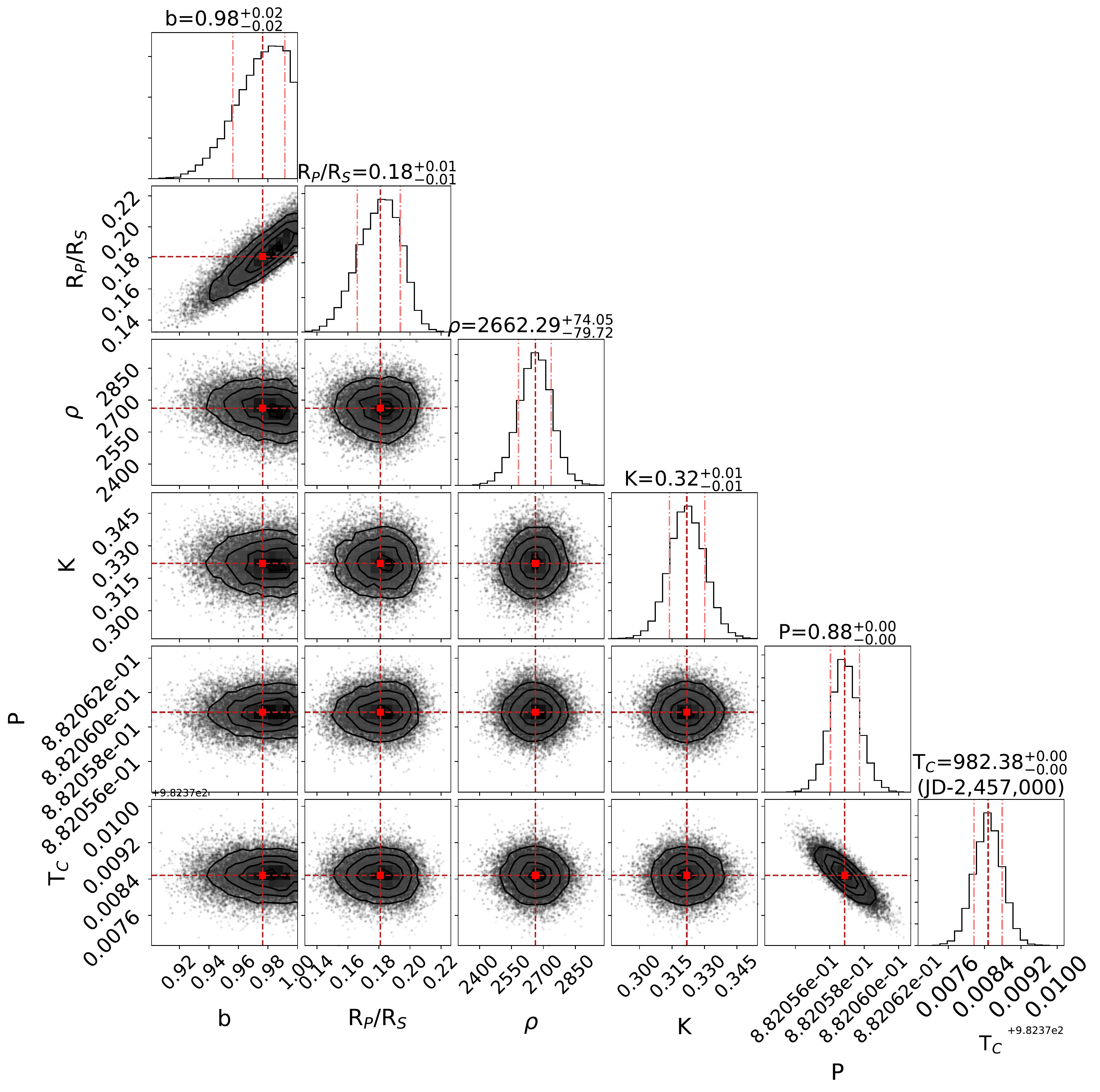}
	\vspace{-5mm}
    \caption{Juliet posterior distributions for the main planetary parameters. The red dashed lines are the median of each distribution and the dash-dot lines represent the $1\sigma$ confidence interval. A correlation between the impact parameter and the planet to star radius is expected due to the grazing nature of the system.}
    \label{fig:juliet_corner}
\end{figure}

The radial-velocity fit made by EMPEROR shows a low eccentricity orbit ($e < 0.01$) thus for the Juliet modeling we decided to fix the eccentricity to 0. Since we have $213,549$ NGTS photometry datapoints plus SAAO photometry in the $V$ and $I$ bands and TESS photometry, fitting such a large light curve is resource intensive, so we first binned the NGTS data in 10 minute cadence bins and then performed the fit with the binned data, supersampling the model light curve to 10 minute exposure times with 30 points in each bin. We also employed supersampling for the 30 minute TESS observations. For the limb darkening we assumed a quadratic law for each instrument. Using Juliet's and the SED fitting output and assuming a Jupiter's Bond albedo of  0.503 \citep{2018NatCo...9.3709L} we calculated the equilibrium temperature of NGTS-6b to be 1283.90$^{+12.49}_{-12.14}$ K. The parameters for the best fit are presented in Table \ref{tab:planet} and in Figure \ref{fig:juliet_corner} we show a corner plot with the main planetary parameters.

The light curves showcased in Section \ref{sec:phot_follow} show a clear V shape transit, suggesting the system is in fact, grazing. This introduces a strong degeneracy between the planet-to-star radius ratio and impact parameter and can produce extreme results (such as an extremely inflated planet). In order to address this issue, a prior for the stellar density constructed from the results of the SED fitting routine was used within Juliet, which allowed us to better decorrelate those two parameters and thus get more realistic results for the parameters of the planet.

\begin{table}
	\centering
	\caption{Planetary Properties for NGTS-6b}
	\begin{tabular}{lc} 
	Property	&	Value \\
	\hline
    P (days)		&	0.882059 $\pm$ 0.0000008\\\vspace{2pt}
	T$_C$ (BJD - 2450000)&	7982.3784 $\pm$ 0.0003	\\\vspace{1pt}
    $a/R_{*}$		& 4.784$^{+0.043}_{-0.048}$\\\vspace{2pt}
    $b$ & 0.976$^{+0.015}_{-0.020}$    \\\vspace{2pt}
	K (\kms) 	& 0.322 $\pm$ 0.008	\\
    e 			& 0.0 {\em (fixed)} 	\\\vspace{2pt}
    \mpl (\mjup)& 1.339 $\pm$ 0.028	\\\vspace{2pt}
    \rpl (\rjup)& 1.326$^{+0.097}_{-0.112}$  \\\vspace{2pt}
    $\rho_{p}$ (\gccc) & 0.711$^{+0.214}_{-0.136}$\\\vspace{2pt}
    a (AU) & 0.01677 $\pm$ 0.00032 \\\vspace{2pt}
    inc (deg)  & 78.231$^{+0.262}_{-0.210}$ \\
    T$_{eq}$ (K)  & 1283.90$^{+12.49}_{-12.14}$ \\
	\hline
	\end{tabular}
    \label{tab:planet}
\end{table}

\begin{figure}
	\includegraphics[width=\columnwidth]{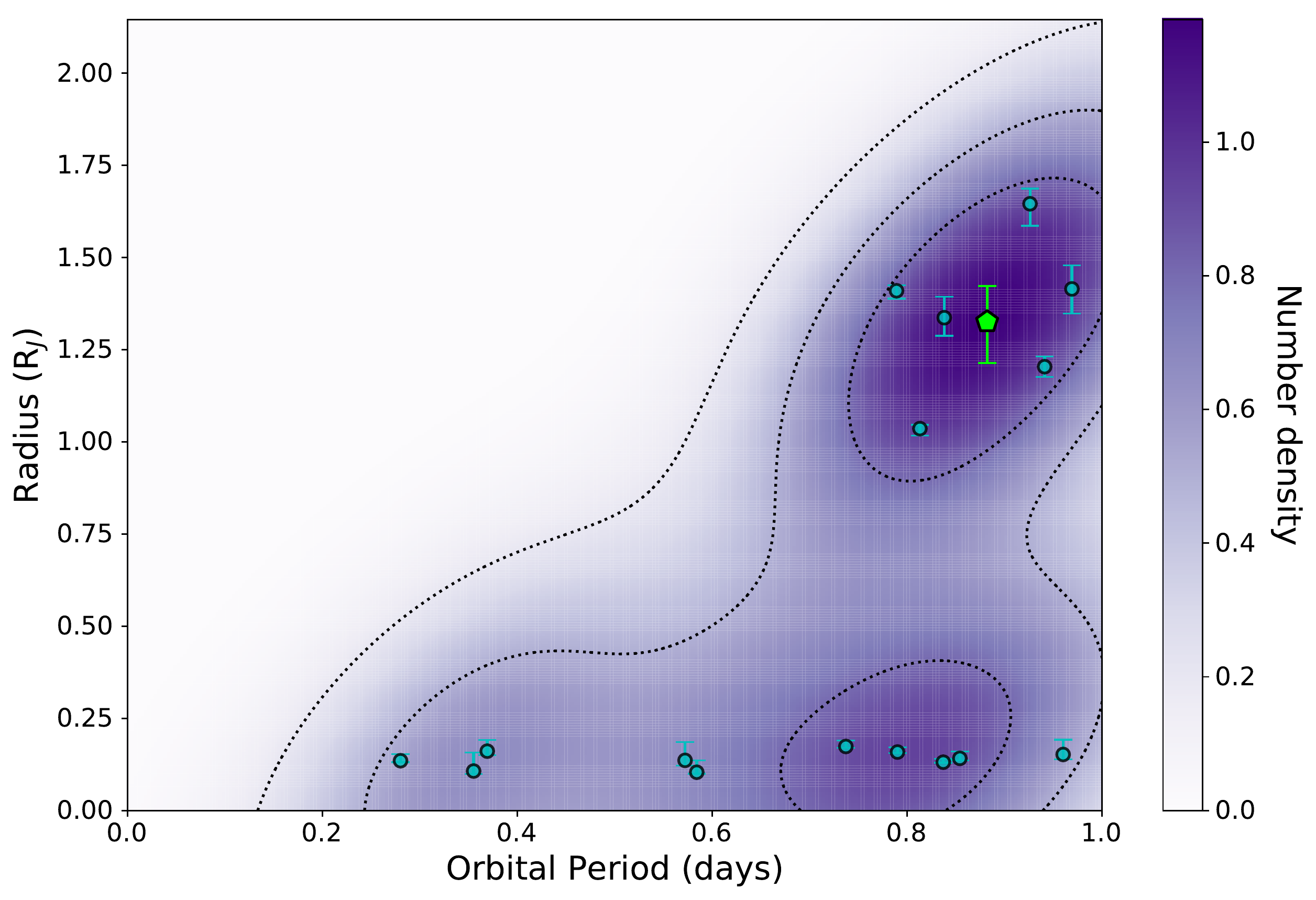}
    \caption{Planetary radius against orbital period. Plotted are all USP planets and UHJs from the well-studied transiting planets catalog that have both measured mass and radius.  The dark contours and purple shading highlight the planet number density of the sample.  The green pentagon shows the position of NGTS-6b}
    \label{fig:rad_period}
\end{figure}

\begin{figure}
	\includegraphics[width=\columnwidth]{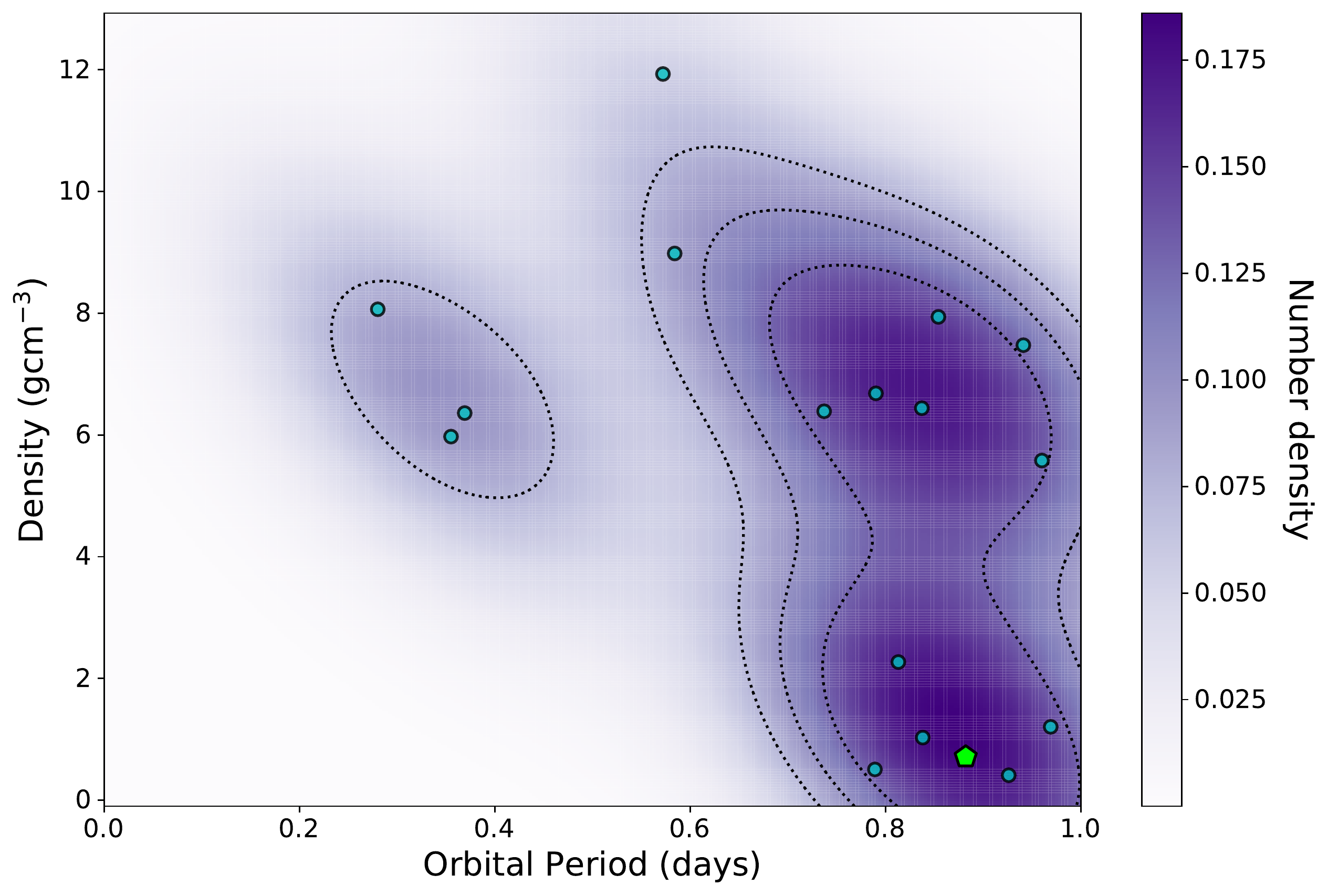}
    \caption{Similar to Figure \ref{fig:rad_period} except we show the planet bulk density against orbital period.}
    \label{fig:den_period}
\end{figure}

\begin{figure}
	\includegraphics[width=\columnwidth]{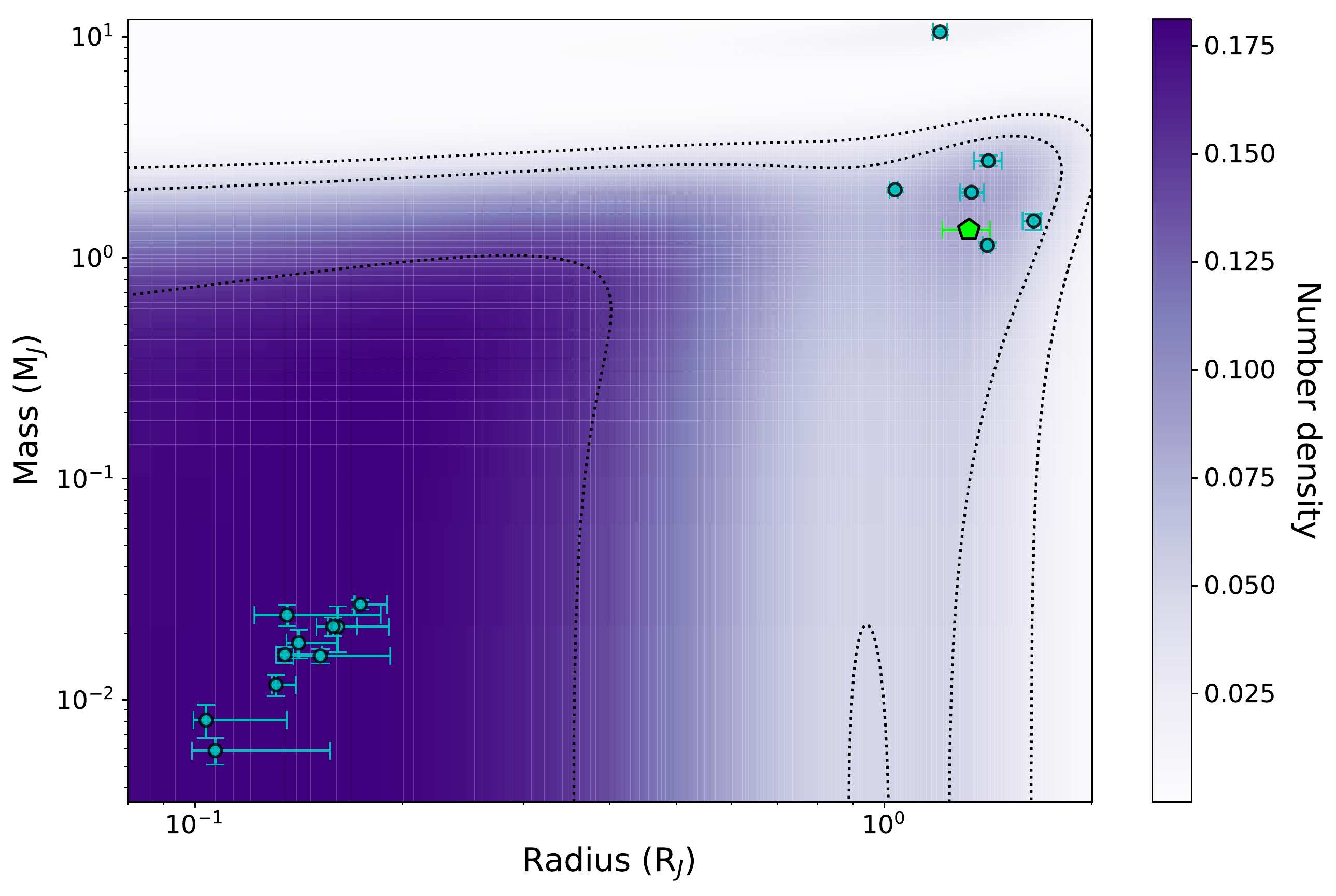}
    \caption{Similar to Figures \ref{fig:rad_period} and \ref{fig:den_period} except here we show the planet mass against planet radius.}
    \label{fig:mass_rad}
\end{figure}

\section{Discussion and Conclusion}
\label{sec:close}

We report the discovery of NGTS-6b, a grazing transit USP HJ with a period of 21.17 hours, mass of  $1.330^{+0.024}_{-0.028}$\mjup\, and radius of $1.271^{+0.197}_{-0.188}$\rjup, and the first USP HJ from the NGTS.  We analyzed the joint photometry and radial velocity data using Juliet, testing its modeling abilities when given a likely grazing transit. There are only a handful of USP planets in the literature, of which only six are giant planets  with \rpl$\,> 8\,$\rearth (WASP-18b, WASP-43b, WASP-103b, HATS-18b, KELT-16b, and WASP-19b) and therefore our discovery of NGTS-6b represents a significant addition to this extreme population. In Figures \ref{fig:rad_period}, \ref{fig:den_period}, \ref{fig:mass_rad} we show that NGTS-6b sits at the centre of the distinct clump of UHJs, and thus adds weight to this being a distinct population.

We investigated photoevaporation of the planet, applying empirical relations from \citet{Jackson2012}, linking the ratio of the X-ray and bolometric luminosities, $L_{\rm X}/L_{\rm bol}$, with stellar age. Using the isochrones-derived age of 9.77\,Gyr yields an estimate of $L_{\rm X}/L_{\rm bol} = 1.0 \times 10^{-5}$ at the current epoch. This corresponds to an X-ray luminosity $L_{\rm X} = 6 \times 10^{27}$\,erg\,s$^{-1}$, or a flux at Earth of $5 \times 10^{-16}$\,erg\,s$^{-1}$\,cm$^{-2}$. Such a flux would require a very deep observation with current generation X-ray telescopes in order to detect the star. Using the energy-limited method of estimating atmospheric mass loss \citep{Watson1981,Erkaev2007}, our estimate of $L_{\rm X}$ yields a mass loss rate of $1\times10^{11}$\,g\,s$^{-1}$. By integrating the mass loss rate across the lifetime of the star \citep[following the X-ray evolution described by][]{Jackson2012} we estimate a total mass loss of about 5 per cent. This is not enough to have significantly evolved the planet, in line with theoretical studies of HJs \citep[e.g.][]{MurrayClay2009,Owen2012}.

We found the host star is likely metal-rich, with a value for the iron abundance [Fe/H] of +0.11$\pm$0.09~dex.  It is well established that gas giant planets favour metal-rich stars \citep{1997MNRAS.285..403G, 2002Msngr.110...32S, 2005ApJ...622.1102F,2015AJ....149...14W}, and also short period gas giants, including the HJ population, also appear even more metal-enhanced when compared to their longer period cousins \citep{2017MNRAS.466..443J, 2018A&A...612A..93M}.  This trend appears to continue into the USP planet population also \citep{Winn_2017}. Approximately 50\% of the USP HJ planets host stars are found to have super-solar metallicities ([Fe/H]$\geq +0.1$~dex), whereas for the smaller super-Earth population, only 30\% orbit such stars.  Since the USP HJ sample is still significantly smaller than the super-Earth sample, NGTS-6b adds statistical weight to this finding, and the conclusion that this points towards is that both these populations form through core accretion processes \citep{2007ApJ...662.1282M}, with the HJ sample forming at relatively large separations from their host stars, and later migrating inwards either through disk driven migration \citep{2007ApJ...660..823M, 2014MNRAS.444.1738T} or high-eccentricity processes like planet-planet scattering \citep{1996Sci...274..954R, 2001Icar..150..303F, 2001MNRAS.325..221P, 2008ApJ...686..621F}.

In \citet{mazeh16} they defined the upper and lower boundaries of the so called "Neptune desert" region, where it was earlier found that there exists a lack of intermediate mass planets \citep[see][]{szabo11,lundkvist16}. Since photoevaporation does not appear to have affected the evolution of NGTS-6b significantly, in line with studies of other HJs that define the upper boundary of this desert \citep{2018A&A...610A..63D}, the planet may have arrived at its current location through high-eccentricity evolution.  \citet{owen2018} suggest that a combination of tidal driven migration to short period orbits through dynamical interactions with other planetary-mass bodies in the system, coupled with photoevaporation of planetary atmospheres can readily describe this sub-Jovian boundary.  

The likelihood of a planet-planet scattering evolutionary scenario for NGTS-6b may also be bolstered if the star is indeed metal-rich.  We can envisage that the planet core quickly grew to a size that crossed the critical core mass limit \citep{mizuno80}, allowing significant accretion of the surrounding gas in the disk.  Yet with a metal-rich proto-planetary disk there would be a high fraction of solids remaining for further planetesimals to form close enough to the young NGTS-6b that they could interact and be scattered to wider orbits, or ejected completely from the system \citep{2018arXiv180405065P}.  Most USP planets are associated with longer period companions \citep{2014ApJ...787...47S, 2017AJ....153...82A, USP_WINN}, where $52\% \pm 5\%$ of HJs have additional, longer period companions \citep{2016ApJ...821...89B}.  A concerted effort to search for additional planets further out in the system, whilst constraining better the orbit of NGTS-6b, may shed some light on these scenarios.

\rm


\section*{Acknowledgements}

Based on data collected under the NGTS project at the ESO La Silla Paranal Observatory.  The NGTS facility is operated by the consortium institutes with support from the UK Science and Technology Facilities Council (STFC)  project ST/M001962/1. 
This paper includes data collected by the TESS mission. Funding for the TESS mission is provided by the NASA Explorer Program. This paper uses observations madeat  the  South  African  Astronomical  Observatory  (SAAO).
PE and AC\\
JIV acknowledges support of CONICYT-PFCHA/Doctorado Nacional-21191829, Chile.
JSJ acknowledges support by Fondecyt grant 1161218 and partial support by CATA-Basal (PB06, CONICYT).  Contributions at the University of Geneva by DB, FB, BC, LM, and SU were carried out within the framework of the National Centre for Competence in Research "PlanetS" supported by the Swiss National Science Foundation (SNSF).
The contributions at the University of Warwick by PJW, RGW, DLP, FF, DA, BTG and TL have been supported by STFC through consolidated grants ST/L000733/1 and ST/P000495/1. 
The contributions at the University of Leicester by MGW and MRB have been supported by STFC through consolidated grant ST/N000757/1.
TL was also supported by STFC studentship 1226157.
MNG is supported by the STFC award reference 1490409 as well as the Isaac Newton Studentship.
EG gratefully acknowledges support from Winton Philanthropies in the form of a Winton Exoplanet Fellowship.
SLC acknolwedges support from an STFC Ernest Rutherford Fellowship.
PE, ACh, and HR acknowledge the support of the DFG priority program SPP 1992 "Exploring the Diversity of Extrasolar Planets" (RA 714/13-1).
This project has received funding from the European Research Council (ERC) under the European Union's Horizon 2020 research and innovation programme (grant agreement No 681601).
The research leading to these results has received funding from the European Research Council under the European Union's Seventh Framework Programme (FP/2007-2013) / ERC Grant Agreement n. 320964 (WDTracer).
We thank Marissa Kotze (SAAO) for developing the SHOC camera  data  reduction  pipeline
We  thank  the  Swiss  National  Science  Foundation  (SNSF) and the Geneva University for their continuous support to our planet search programs. This work has been in particular carried out in the frame of the National Centre for Competence in Research PlanetS supported by the Swiss National Science Foundation (SNSF).
This publication makes use of The Data \& Analysis Center for Exoplanets (DACE), which is a facility based at the University of Geneva (CH) dedicated to extrasolar planets data visualisation, exchange and analysis. DACE is a platform of the Swiss National Centre of Competence in Research (NCCR) PlanetS, federating the Swiss expertise in Exoplanet research. The DACE platform is available at \url{https://dace.unige.ch}. 
The Pan-STARRS1 Surveys (PS1) and the PS1 public science archive have been made possible through contributions by the Institute for Astronomy, the University of Hawaii, the Pan-STARRS Project Office, the Max-Planck Society and its participating institutes, the Max Planck Institute for Astronomy, Heidelberg and the Max Planck Institute for Extraterrestrial Physics, Garching, The Johns Hopkins University, Durham University, the University of Edinburgh, the Queen's University Belfast, the Harvard-Smithsonian Center for Astrophysics, the Las Cumbres Observatory Global Telescope Network Incorporated, the National Central University of Taiwan, the Space Telescope Science Institute, the National Aeronautics and Space Administration under Grant No. NNX08AR22G issued through the Planetary Science Division of the NASA Science Mission Directorate, the National Science Foundation Grant No. AST-1238877, the University of Maryland, Eotvos Lorand University (ELTE), the Los Alamos National Laboratory, and the Gordon and Betty Moore Foundation.



\bibliographystyle{mnras}
\bibliography{ref.bib} 








\bsp	
\label{lastpage}
\end{document}